\documentclass{aa}  
\bibpunct{(}{)}{;}{a}{}{,}
\usepackage[singlelinecheck=off]{caption}
\usepackage{graphicx}
\usepackage{longtable}
\usepackage{leftidx}
\usepackage{amssymb}
\usepackage{gensymb}
\usepackage{textcomp}
\usepackage{epstopdf}

\newcommand{\asec}{$^{\prime\prime}$}
\usepackage{txfonts}

\begin{document}

   \title{The trans/cis ratio of formic (HCOOH) and thioformic (HC(O)SH) acids in the interstellar medium}

   \author{J. García de la Concepción
          \inst{1}
          \and
          L. Colzi
          \inst{1,2}
          \and
          I. Jiménez-Serra
          \inst{1}
          \and
          G. Molpeceres
          \inst{3}
          \and
          J. C. Corchado
          \inst{4}
          \and
          V. M. Rivilla
          \inst{1,2}
          \and
          J. Martín-Pintado
          \inst{1}
          \and
          M. T. Beltrán
          \inst{2}
          \and
          C. Mininni
          \inst{2,5,6}
          }

   \institute{Centro de Astrobiología (CSIC-INTA), Ctra. de    Ajalvir Km. 4, Torrejón de Ardoz, 28850 Madrid, Spain\\
              \email{jgarcia@cab.inta-csic.es}
              \and
         INAF-Osservatorio Astrofisico di Arcetri, Largo E. Fermi 5, I-50125, Florence, Italy
         \and
         Institute for Theoretical Chemistry, University of Stuttgart, Pfaffenwaldring 55, 70569 Stuttgart, Germany
         \and 
         Departamento de Ingeniería Química y Química Física, Facultad de Ciencias, and ICCAEx, Universidad Extremadura, Badajoz, Spain
         \and
         Università degli studi di Firenze, Dipartimento di fisica e Astronomia, Via Sansone 1, 50019 Sesto Fiorentino, Italy
         \and
         INAF-IAPS, via del Fosso del Cavaliere 100, I-00133 Roma, Italy}

   \date{}

 
  \abstract
   {Observations of the different isomers of molecules in the interstellar medium (ISM) have revealed that both low- and high-energy isomers can be present in space despite the low temperature conditions. It has been shown that the presence of these isomers may be due to tunneling effects.}
   {We carried out a theoretical study of the cis--trans isomerization reactions of two astrophysically relevant acids, formic acid (HCOOH) and thioformic acid (HC(O)SH), where the latter has recently been discovered in space. We also searched for these molecules towards the hot core G31.41+0.31 to compare their abundances with the expected theoretical isomerization results.}
   {We employed high-level $ab$ $initio$ calculations to study the reaction rate constants of the isomerization reactions. We used the canonical variational transition state theory (CVT) with the multidimensional small curvature tunneling (SCT) approximation in the temperature range of 10-400 K. Moreover, we used the spectrum obtained from the ALMA 3mm spectral survey GUAPOS (GUAPOS: G31 Unbiased ALMA sPectral Observational Survey), with a spectral resolution of $\sim$0.488 MHz and an angular resolution of 1\farcs2$\times$1\farcs2 ($\sim$4500 au), to derive column densities of HCOOH and HC(O)SH towards G31.41+0.31.}
   {Our results demonstrate that these isomerizations are viable in the conditions of the ISM due to ground-state tunneling effects, which allow the system to reach the thermodynamic equilibrium at moderately low temperatures. At very low temperatures (T$_{kin}$ $\sim$10 K), the reaction rate constants for the cis-to-trans isomerizations are very small, which implies that the cis isomers should not be formed under cold ISM conditions. This is in disagreement with observations of the cis/trans isomers of HCOOH in cold cores where the cis isomer is found to be $\sim$5-6\% the trans isomer. At high temperatures ($\sim$150-300 K), our theoretical data not only match the observed behavior of the trans/cis abundance ratios for HCOOH (the cis form is undetected), but they support our tentative detection of the trans and ---for the first time in the insterstellar medium--- the cis isomer of HC(O)SH towards the hot molecular core G31.41+0.31 (with a measured trans/cis abundance ratio of $\sim$3.7).}
   {While the trans/cis ratio for HC(O)SH in the ISM depends on the relative stability of the isomers, the trans/cis ratio for HCOOH cannot be explained by isomerization, and is determined by other competitive chemical processes.}

                
\titlerunning{The trans/cis ratio of formic and thioformic acids in the Interstellar Medium}
   \maketitle

\section{Introduction}

The detection of isomers in the interstellar medium (ISM) has triggered efforts in astrophysics to gain insight into the formation path of these species. As an illustration, let us consider the different values of the E/Z isomer ratios of imines such as cyanomethanimine, ethanamine, and acetylenimine measured across different environments of the ISM \citep{loomis13,victor19,bizzocchi20}. Several theories have been proposed to explain the different  observed E/Z ratios. The most popular approach considers that the different isomers should be formed through competitive chemical routes that selectively lead to the observed ratio \citep{puzzarini15,quan16,balucani18,melli18,baiano20,bizzocchi20,lupi20,Shingledecker20}. This approach implicitly assumes that the thermodynamic equilibrium between the isomers cannot be reached under the conditions of the ISM due to the high energy barriers of the isomerization reactions ($\sim$25-30 kcal/mol, \citealt{juan,vazart15,balucani18,baiano20}). However, in previous works,  the quantum tunneling corrections to the rate coefficients were carried out using a one-dimensional treatment (see section 2.3. in \citealt{bao}). More recently, we demonstrated that multidimensional treatment of this quantum effect (e.g., where all the degrees of freedom of the molecule are considered in the tunneling process) increases the transmission coefficients by several orders of magnitude \citep{juan}. As a result, the ratio between the E/Z isomers of cyanomethanimine, ethanamine, and propinimine depends on the isomer relative stability because the thermodynamic equilibrium can be reached in the ISM thanks to ground-state quantum tunneling effects \citep{juan}.

Recently, the trans isomer of thioformic acid, t-HC(O)SH, was discovered in the ISM \citep[with an abundance of 1.2$\times$10$^{-10}$;][]{lucas21}. This species was found toward the giant molecular cloud G+0.693-0.027 located in the Galactic center (with a kinetic temperature of T$_{kin}$ \(\sim\)150 K), where only the trans isomer could be detected \citep[the derived upper limit abundance of the cis isomer, c-HC(O)SH, is $\leq$0.2$\times$10$^{-10}$;][]{lucas21}. The selective formation of t-HC(O)SH on dust grains has been found to be very efficient through the hydrogenation of OCS in ices \citep{germanjuan}. However, this is the only detection of this species in the ISM so far \citep{lucas21}, and it remains unknown whether it could be present toward other sources, either in its trans isomeric form or in both trans and cis forms.

Likewise, the trans isomer of the formic acid, t-HCOOH, has been widely detected in the ISM, in cold pre-stellar cores such as L1544 or L183 (\citealt{vastel2014}; \citealt{lattanzi2020}) and low-mass protostars (e.g., \citealt{yang2021}; \citealt{maureira2020}; \citealt{manigand2020}; \citealt{vangelder2020}), to protoplanetary disks (\citealt{favre2018}; \citealt{lee2019}) and massive star-forming regions both in the Galactic disk (e.g., \citealt{rivilla2017a} \citealt{gieser2019}; \citealt{peng2019}) and in the Galactic center (e.g., \citealt{belloche2013}; \citealt{juan2020}; \citealt{lucas21}). Unlike t-HCOOH, the cis isomer, c-HCOOH has been only found towards sources such as the photon-dominated region (PDR) of
the  Orion
Bar \citep[with a trans/cis abundance ratio of 2.8;][]{cuadrado16}, and the dense cold molecular clouds (T$_{kin}$ \(\sim\)10 K) L483 \citep{agundez19} and B5 \citep{taquet17}, with a trans/cis ratio of \(\sim\)16.7. 
For the Orion Bar PDR, high-energy UV photons excite the molecule of HCOOH to the first singlet excited state, and then the deactivation of the excited state to the ground state of the cis and trans isomers enables the observations of both conformers. For the cold dense clouds, there are no incident UV photons, but there are secondary UV fields produced by the relaxation of hydrogen from high excited states that can be generated by cosmic rays. This latter process could also alter the trans/cis isomer ratio. This questions whether the observed trans/cis ratios are due to a temperature effect that would allow the formation of the cis isomer, or whether the systems are in thermodynamic equilibrium as observed for imines. The prevalence of one isomer over the other as a function of the T$_{kin}$ could be of great importance to better understand the chemistry of the ISM.

It is well known that simple carboxylic acids like HCOOH acid can isomerize from the least stable isomer (cis) to the most stable (trans) through quantum tunneling at very low temperatures \citep{pettersson16,macoas05,domanskaya09,tsuge14}. However, this effect has not yet been explored for thiocarboxylic acids, like HC(O)SH. In this study, we carried out a theoretical analysis of the isomerization reaction of the HCOOH and HC(O)SH acids at interstellar temperatures (T$_{kin}$=10-400 K) in vacuum, with the aim of understanding whether or not the prevalence of one isomer over the other in the ISM is due to an isomerization transformation. To account for the quantum tunneling effect, which is very important especially at low temperatures, we selected the multidimensional small curvature tunneling (SCT) approximation, which has been demonstrated to give much higher transmission coefficients than other conventional monodimensional treatments \citep{juan}.

In section 2, we describe the computational details used in this work and in  Section 3 we explain the results obtained, both the electronic structure calculations and the kinetic results. In Section 4, we present the detection of t-HCOOH and the tentative detection of  c-HC(O)SH and t-HC(O)SH toward the massive hot core G31.41+0.31 (G31 hereafter). In Section 5, we compare the trans/cis ratios found toward several astronomical sources with our theoretical predictions. In Section 6 we give the conclusions of this work.

\section{Computational details}
\subsection{Electronic structure calculations}

All the stationary points were optimized without constraints at the double-hybrid B2PLYP \citep{grimme06} with the dispersion-corrected D3(BJ) \citep{grimme10,grimme11} method in combination with the Dunning's triple-zeta correlation-consistent basis set, augmented with diffuse functions for all atoms (aug-cc-pVTZ) \citep{dunning89,kendall92}. The same level of theory was used to compute frequency calculations in order to characterize the stationary points, which showed none and one imaginary frequency for the energy minima and saddle points, respectively. To improve the level of theory, we corrected the electronic energies of the optimized geometries at B2PLYP-D3(BJ)/aug-cc-pVTZ with the CCSD(T)-F12 method \citep{adler2007,knizia2009} which includes correlated basis functions and the RI approach for the correlation integrals \citep{feyereisen93}. As auxiliary and special orbital basis set, the cc-pVTZ-F12-CABS and cc-pVTZ-F12 were used in combination with the augmented quadruple zeta aug-cc-pVQZ for the correlation and coulomb fitting. Finally, we corrected the electronic energies by computing the anharmonic zero-point energy (ZPE) of the stationary points within vibrational perturbative theory to second order \citep{Barone2004}. The relative energies are given with respect to the least stable isomer (cis). In order to validate the choice of the double-hybrid method we also optimized all the stationary points for HCOOH with the CCSD(T)-F12 method in conjunction with the cc-pVTZ-F12-CABS and cc-pVTZ-F12 basis set and the aug-cc-pVTZ for the correlation and coulomb fitting (see Appendix A). We also computed the frequencies at the same level of theory, finding one and none imaginary frequencies for the saddle point and energy minima respectively. The comparison between these two methods shows that the B2PLYP-D3(BJ) is suitable for quantitative investigations, because both the relative energies and the geometries of the stationary points are almost identical (see Tables A.1 and A.2 and Figure A.1). The second-order perturbation theory was used to study the orbital interactions. For the geometry optimization with the double-hybrid functional, and the correction of the electronic energies with the coupled-cluster method, the Gaussian 16 \citep{gaussian} and ORCA \citep{neese12,neese20} software packages were employed respectively. For visualization of the geometries and the orbital isosurfaces we used the Cylview \citep{cylview} and Chimera \citep{chimera} softwares respectively.

\subsection{Kinetics calculations}

We used the canonical variational transition state theory (CVT) \citep{bao} to compute rate constants with the small curvature tunneling approximation (SCT) in the temperature ($T$) range of 10-400 K. The expression for the unimolecular rate constants is given by

\begin{equation}
K^{CVT/SCT}(T) = \kappa^{SCT} \Gamma^{CVT} \frac{K_{B}T}{h} \frac{Q^{VT}}{Q^{R}} e(-E^{VT}/{RT}) 
,\end{equation}

where $R$ is the ideal gas constant; \(\kappa^{SCT}\) is the small curvature multidimensional tunneling transmission coefficient; \(\Gamma^{CVT}\) is the canonical variational transition state recrossing coefficient, which is given by \(\textit{K}\)$^{CVT}$/\(\textit{K}\)$^{TST}$, where \(\textit{K}\)$^{TST}$ is the rate coefficient of the conventional transition state theory, whereas \(\textit{K}\)$^{CVT}$ is the rate coefficient of the canonical variational transition state theory; K$_{B}$ is the Boltzmann constant and $h$ is the Planck constant; $Q^{VT}$ and $Q^{R}$ are the total partition functions of the variational transition state and the reactant, respectively; and \(\textit{E}\)$^{VT}$ is the potential energy of the variational transition state. 
The energies of the minimum energy path (MEP) obtained with the double-hybrid B2PLYP-D3(BJ)/aug-cc-pVTZ were corrected with the coupled cluster method mentioned in section 2.1 using the interpolated single-point energies (ISPE) algorithm \citep{chuang99}. The rate coefficient calculations were carried out with the Pilgrim software \citep{pilgrim}.

Finally, we note that `forward pathway' (i.e., from S $<$ 0 to S $>$ 0, with S being the reaction coordinate in Bohr) hereafter refers to the movement along the reaction path from the cis isomer to the trans isomer. For simplicity, the forward cis to trans rate constants are denoted as $K{_1}$ and the backward trans to cis as $K_{-1}$.

\section{Results}
\label{results} 
\subsection{Stationary points and stereoelectronic effects}

The isomerization reaction of the HC(O)SH involves the rotation of the O-C-S-H dihedral angle in the same way as the O-C-O-H angle for HCOOH. In the cis isomers, the dihedral angle is 180.0$^\circ$, whereas for the trans isomers it is 0.0$^\circ$. Both are connected through a transition structure (TS) with a dihedral angle of 89.6$^\circ$ and 94.3$^\circ$ for HC(O)SH and HCOOH, respectively (see Figure 1).

\begin{figure*}[h!]
  \centering
  \includegraphics[width=14cm,height=11cm]{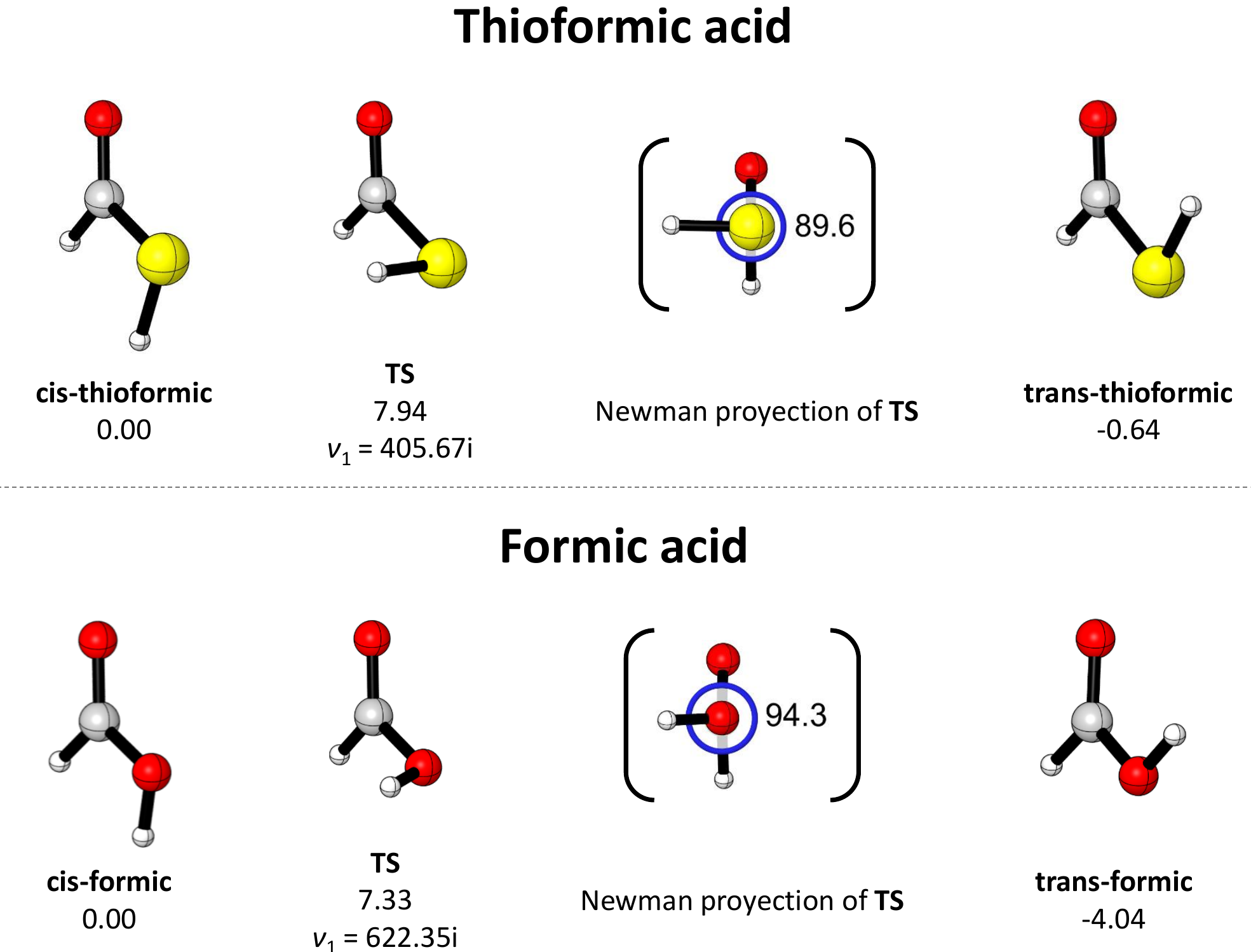}
  \caption{Optimized structures of the cis, trans, and TS of HC(O)SH and HCOOH. Relative energies with respect the cis isomer calculated with the CCSD(T)-F12 method on the B2PLYP-D3(BJ) geometries are given, including anharmonic zero-point energies computed with the B2PLYP-D3(BJ) method.}
\end{figure*}

The values of these degrees of freedom in the transition structures indicate that the geometry of the TS for the HC(O)SH acid is slightly closer to that of the t-HC(O)SH isomer, whereas the TS of the HCOOH acid is closer to that of the c-HCOOH isomer. The energy barrier (with inclusive anharmonic ZPE) relative to the cis isomer is 7.94 and 7.33 kcal/mol for HC(O)SH and HCOOH, respectively. Both energy barriers are  similar. However, the imaginary frequency of the saddle points associated with the reaction coordinate is 216.7 cm$^{-1}$ higher for HCOOH. The latter indicates that the mass of the atom that holds the acidic hydrogen is of relevance in the isomerization reaction.

The relative electronic energy between the two isomers of HC(O)SH is 0.68 kcal/mol, which is notably smaller than that found for HCOOH (4.04 kcal/mol). These energy differences are due to the dipolar moments of the isomers. For instance, the difference between the dipolar moments of the isomers are 1.39 and 2.39 Debye for HC(O)SH and HCOOH respectively, with the most polar being the cis isomers. To easily visualize this, we built the electrostatic potential (ESP) maps of the isomers for both compounds (Figure 2 a)). The polarization of the X-H bond (X being S for HC(O)SH and O for the HCOOH) has an important influence on the polarity of these molecules. The electrostatic potential (ESP) maps of the c-HC(O)SH and t-HC(O)SH isomers are very similar. However, for HCOOH, the differences in the charge distribution between cis and trans isomers are noticeable. This high charge polarization for c-HCOOH makes it unstable under the low pressure conditions of the ISM, because there is not any molecule close enough to this dipole to interact with and stabilize it. The latter statement has been checked by optimizing c-HCOOH and t-HCOOH in water using the continuum solvation model SMD \citep{marenich091,marenich092}. The derived electronic energy difference between the isomers is 0.5 kcal/mol. Although these interactions seem to be determinant in the relative stability of the isomers, the energy differences are also related to the delocalization energies. The second-order perturbative analysis for the cis and trans isomers of both acids shows that the strongest orbital interaction is between the lone pair of the sulfur and oxygen atoms that hold the hydrogen (LP orbital) with the \(\pi\) antibonding of the C-O (\(\pi\)*$_{CO}$) (see Figure 2 b)). In both cases, the highest delocalization energy is for the trans isomers, especially for HCOOH.

\subsection{Kinetic results of the isomerizations}

The analysis of the vibrational frequencies along the MEP for these isomerization transformations shows that variational effects should be negligible because the frequencies remain almost constant along the MEP. This effect is reflected in the relation \(\textit{K}\)$^{CVT}$/\(\textit{K}\)$^{TST}$, which is equal to 1 over the whole range of temperatures between 10 and 400 K. 

An important factor limiting the accuracy of the theoretically computed rate constants at low temperatures is the high-frequency anharmonicities \citep{zheng14,gao18,wu20}. We take the high-frequency anharmonicities into account by multiplying the ZPE by a scaling factor for the energy minima and along the whole reaction path (0.948 for HCOOH and 0.949 for HC(O)SH). This approximation is explained elsewhere \citep{gao18}. 

The imaginary frequency associated with the reaction coordinate for the TS-HC(O)SH is 405.67i, whereas for TS-HCOOH it is 622.35i. This difference in the imaginary frequency can be visualized easily through the vibrationally adiabatic potentials (V$_{a}^{G}$) depicted in Figure 3: V$_{a}^{G}$ is noticeably wider for HC(O)SH. The representative tunneling energies (RTEs) lie above the vibrational ground state (\(\nu\)=0) of the cis isomers. The RTEs remain constant up to 60 K for HC(O)SH and 110 K for HCOOH due to the fact that the V$_{a}^{G}$ potential is markedly narrower for HCOOH, and therefore the transmission probabilities are higher. These differences in V$_{a}^{G}$ influence the cis to trans rate constants, $K{_1}$, especially at 10 K, where they are 3.73x10$^{-21}$ for HCOOH and 6.04x10$^{-26}$ for HC(O)SH. These rate constants suggest that this transformation is unlikely in the ISM at this low temperature.

\begin{figure*}[h!]
  \centering
  \includegraphics[width=15cm,height=22cm]{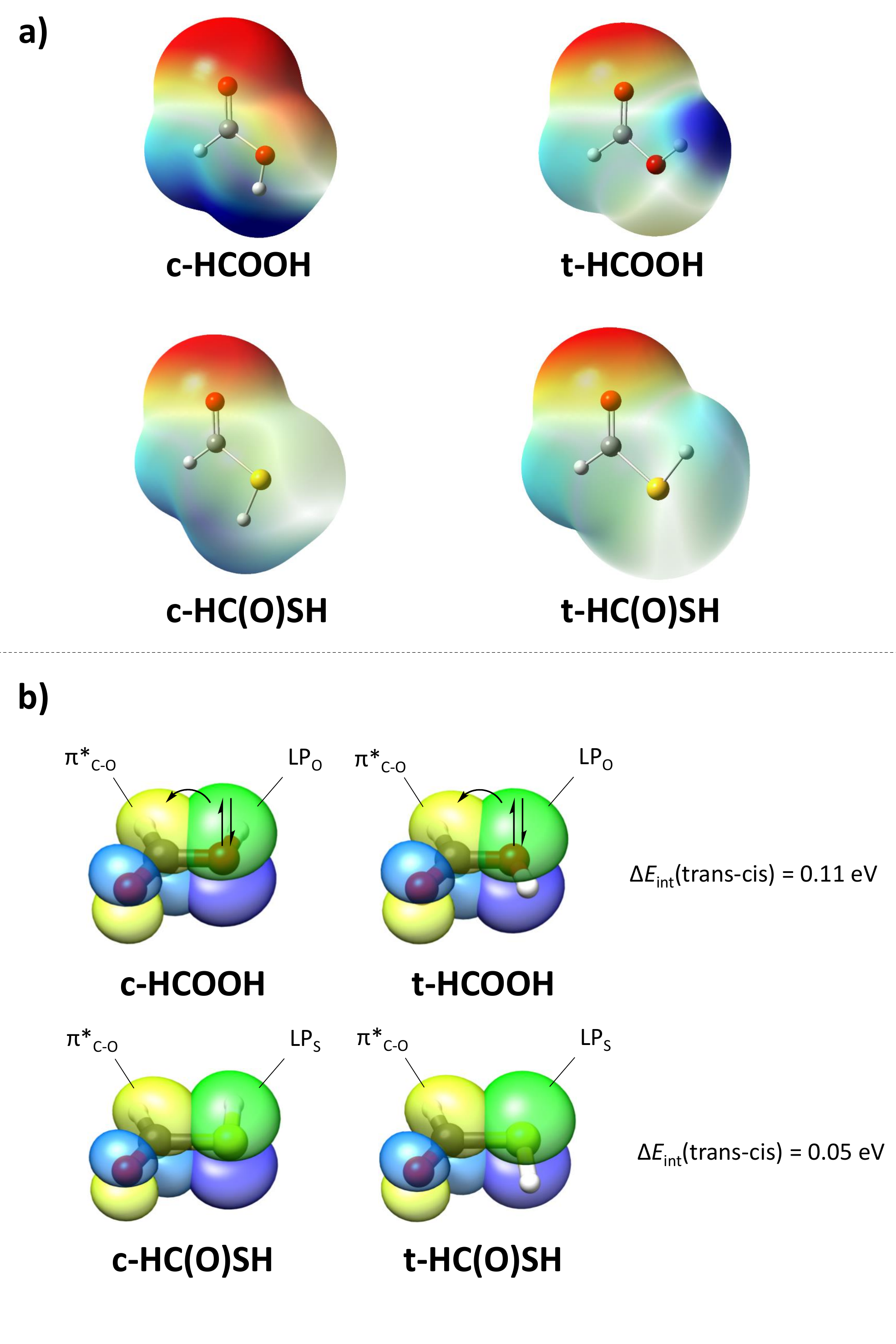}
  \caption{Representation of isosurfaces for the acids HCOOH and HC(O)SH. (a) ESP isosurfaces of both isomers of HCOOH and HC(O)SH with isovalues between -0.04 and 0.05 au. The red and blue areas of the ESP maps represent the highest and lowest electron densities. (b) Isosurfaces of the \(\pi\)*$_{CO}$ and LP$_{S}$ and LP$_{O}$ orbitals for both isomers and interaction energy between these two orbitals due to delocalization.}
\end{figure*}

\begin{figure*}[h!]
  \centering
  \includegraphics[width=15cm,height=20cm]{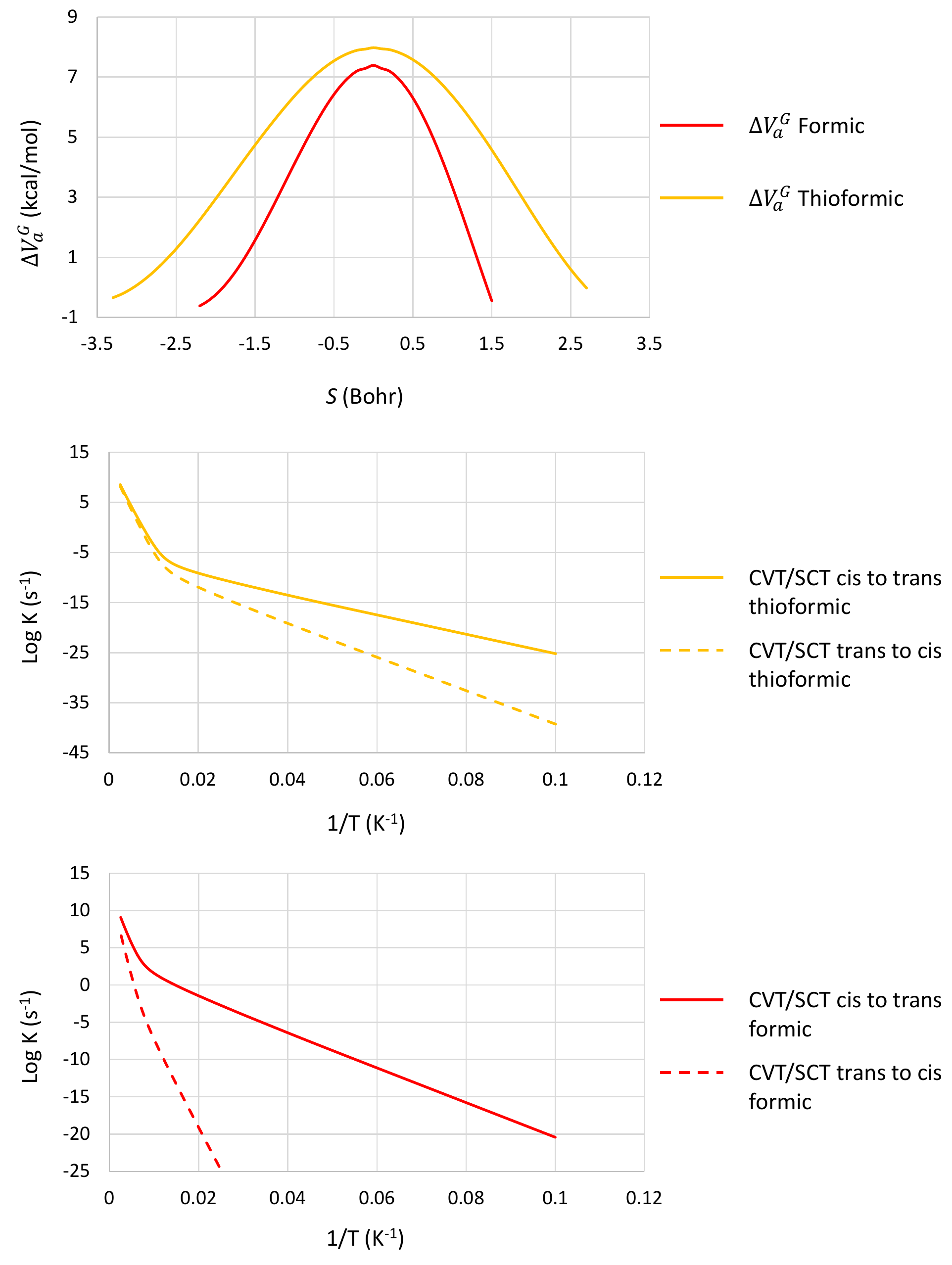}
  \caption{Relative V$_{a}^{G}$ potentials with respect the cis isomers in kcal/mol. Arrhenius plots for the cis to trans (solid lines) and trans to cis (dotted lines) isomerization reactions calculated with the CVT/SCT theory.}
\end{figure*}

\begin{table*}
\setlength{\tabcolsep}{1pt} 
\caption{CVT/SCT rate constants (in s$^{-1}$) for the cis to trans and trans to cis isomerization reactions of thioformic [HC(O)SH] and formic acids [HC(O)OH] and trans/cis ratios.}             
\label{tab1}      
\centering           
\begin{tabular}{c c c c c c c}  
\hline\hline    
T (K) & \multicolumn{3}{c}{Thioformic acid [HC(O)SH]} & \multicolumn{3}{c}{Formic acid [HC(O)OH]}\\
\hline
 & $K^{CVT/SCT}$ [cis to trans] & $K^{CVT/SCT}$ [trans to cis] & $K{_1}$/$K_{-1}$ & $K^{CVT/SCT}$ [cis to trans] & $K^{CVT/SCT}$ [trans to cis] & $K{_1}$/$K_{-1}$\\
\hline                        
10      &       6.04E-26        &       5.14E-40        &       1.17E+14        &       3.73E-21        &       1.79E-109       &       2.09E+88        \\
20      &       3.03E-16        &       2.78E-23        &       1.09E+07        &       1.67E-09        &       1.14E-53        &       1.47E+44        \\
30      &       7.34E-13        &       1.49E-17        &       4.93E+04        &       1.64E-05        &       5.82E-35        &       2.82E+29        \\
40      &       4.74E-11        &       1.43E-14        &       3.32E+03        &       1.89E-03        &       1.53E-25        &       1.23E+22        \\
50      &       7.65E-10        &       1.17E-12        &       6.57E+02        &       3.65E-02        &       7.70E-20        &       4.74E+17        \\
60      &       7.04E-09        &       3.16E-11        &       2.23E+02        &       2.89E-01        &       5.36E-16        &       5.40E+14        \\
70      &       6.04E-08        &       5.85E-10        &       1.03E+02        &       1.39E+00        &       3.26E-13        &       4.26E+12        \\
80      &       7.31E-07        &       1.26E-08        &       5.78E+01        &       4.93E+00        &       4.37E-11        &       1.13E+11        \\
90      &       1.44E-05        &       3.90E-07        &       3.69E+01        &       1.46E+01        &       2.18E-09        &       6.70E+09        \\
100     &       3.23E-04        &       1.26E-05        &       2.57E+01        &       3.88E+01        &       5.54E-08        &       7.00E+08        \\
110     &       5.90E-03        &       3.09E-04        &       1.91E+01        &       9.82E+01        &       8.91E-07        &       1.10E+08        \\
120     &       7.81E-02        &       5.23E-03        &       1.49E+01        &       2.47E+02        &       1.05E-05        &       2.36E+07        \\
130     &       7.55E-01        &       6.24E-02        &       1.21E+01        &       6.36E+02        &       9.93E-05        &       6.41E+06        \\
140     &       5.53E+00        &       5.47E-01        &       1.01E+01        &       1.68E+03        &       8.00E-04        &       2.09E+06        \\
150     &       3.20E+01        &       3.71E+00        &       8.64E+00        &       4.45E+03        &       5.60E-03        &       7.94E+05        \\
160     &       1.52E+02        &       2.02E+01        &       7.53E+00        &       1.16E+04        &       3.42E-02        &       3.40E+05        \\
170     &       6.08E+02        &       9.12E+01        &       6.67E+00        &       2.93E+04        &       1.82E-01        &       1.61E+05        \\
180     &       2.11E+03        &       3.52E+02        &       5.99E+00        &       7.02E+04        &       8.50E-01        &       8.26E+04        \\
190     &       6.47E+03        &       1.19E+03        &       5.43E+00        &       1.60E+05        &       3.51E+00        &       4.55E+04        \\
200     &       1.79E+04        &       3.59E+03        &       4.97E+00        &       3.43E+05        &       1.29E+01        &       2.66E+04        \\
220     &       1.05E+05        &       2.45E+04        &       4.27E+00        &       1.36E+06        &       1.29E+02        &       1.05E+04        \\
240     &       4.63E+05        &       1.23E+05        &       3.77E+00        &       4.47E+06        &       9.23E+02        &       4.84E+03        \\
260     &       1.64E+06        &       4.86E+05        &       3.38E+00        &       1.26E+07        &       5.02E+03        &       2.51E+03        \\
280     &       4.90E+06        &       1.59E+06        &       3.08E+00        &       3.12E+07        &       2.18E+04        &       1.43E+03        \\
300     &       1.27E+07        &       4.47E+06        &       2.84E+00        &       6.94E+07        &       7.92E+04        &       8.76E+02        \\
320     &       2.94E+07        &       1.11E+07        &       2.65E+00        &       1.41E+08        &       2.47E+05        &       5.70E+02        \\
340     &       6.18E+07        &       2.48E+07        &       2.49E+00        &       2.66E+08        &       6.81E+05        &       3.91E+02        \\
360     &       1.20E+08        &       5.09E+07        &       2.35E+00        &       4.70E+08        &       1.69E+06        &       2.79E+02        \\
380     &       2.18E+08        &       9.71E+07        &       2.24E+00        &       7.87E+08        &       3.82E+06        &       2.06E+02        \\
400     &       3.72E+08        &       1.74E+08        &       2.14E+00        &       1.26E+09        &       8.00E+06        &       1.57E+02        \\ 
\hline                                   
\end{tabular}
\end{table*}

When the RTE leaves the \(\nu\) = 0 of the cis isomer, there is a transition from the tunneling-controlled reactions to the thermally activated reactions, which can be seen in Figure 3 as a change of the slope of the Arrhenius plots. There is a contribution from quantum tunneling up to approximately 110 K for HC(O)SH and 280 K for HCOOH. The contribution of tunneling to higher temperatures for HCOOH acid with respect to HC(O)SH is also a consequence of the  V$_{a}^{G}$ width.

From Figure 3, we also find that the backward isomerization reactions (from trans to cis) are very different for the two species at very low temperatures (see rate constant values in Table \ref{tab1}). As the RTE lies in the ground state of the cis isomer, the trans conformer of HC(O)SH has to overcome an energy gap of \(\sim\)0.64 kcal/mol, whereas the gap is of \(\sim\)4.04 kcal/mol for HCOOH. This means that the ratio between the forward and backward rate constants is 2.09x10$^{88}$ for HCOOH and 1.17x10$^{14}$ for HC(O)SH at 10 K. Even so, the backward isomerizations can be viable at high temperatures for both acids thanks to ground-state quantum tunneling. The contribution of tunneling to the rate constant in the backward isomerizations is noticeable up to the same temperatures as those for the forward paths.

\section{Comparison with observations from the GUAPOS spectral survey}
\label{observation}

The results presented in the previous section suggest that we should not expect to detect the cis isomer of HCOOH due to isomerization in the ISM because of the large ratios between the rate constants of the forward (cis to trans) and backward (trans to cis) reactions for this acid {at all temperatures}. On the other hand, for HC(O)SH, this would only be true at low temperatures, because the ratio between the rate constants of the forward (cis to trans) and backward (trans to cis) reactions becomes $\leq$10 for temperatures $\geq$150$\,$K. In this section, we analyze the data from the GUAPOS spectral survey \citep{mininni2020}, and compare them to data found in the literature in order to investigate whether the trans/cis abundance ratios of HCOOH and HC(O)SH follow the values found for tunneling and are consistent with the expected ones under thermodynamic equilibrium, as proposed in Section$\,$\ref{results}.

We search for the trans and cis forms of both HCOOH and HC(O)SH acid towards the hot molecular core (HMC) G31 using data from the GUAPOS spectral survey \citep{mininni2020,laura21}. G31 is located at a distance of 3.75 kpc (\citealt{immer2019}), with a luminosity of $\sim$4.5$\times$10$^{4}$ L$_{\odot}$ (from \citealt{osorio2009}) and a mass of $\sim$70 M$_{\odot}$ (\citealt{cesaroni2019}). It harbors at least four massive star-forming cores towards its center (\citealt{beltran2021}). Moreover, different molecular lines present an inverse P-Cygni profile, indicating that the core is collapsing (e.g., \citealt{girart2009}; \citealt{beltran2018}).  The core is also rotating as indicated by a clear NE--SW velocity gradient observed in different high-density tracers (e.g., \citealt{beltran2004}; \citealt{beltran2018}). Towards G31, many molecules have been detected (\citealt{beltran2005}; \citealt{beltran2009};
\citealt{rivilla2017b}; \citealt{beltran2018}; \citealt{mininni2020}; \citealt{gorai2021};
\citealt{laura21}), some of which are complex (>5 atoms). Therefore, G31 is a promising target to search for HCOOH and HC(O)SH.

\begin{figure*}
\centering
\includegraphics[width=35pc]{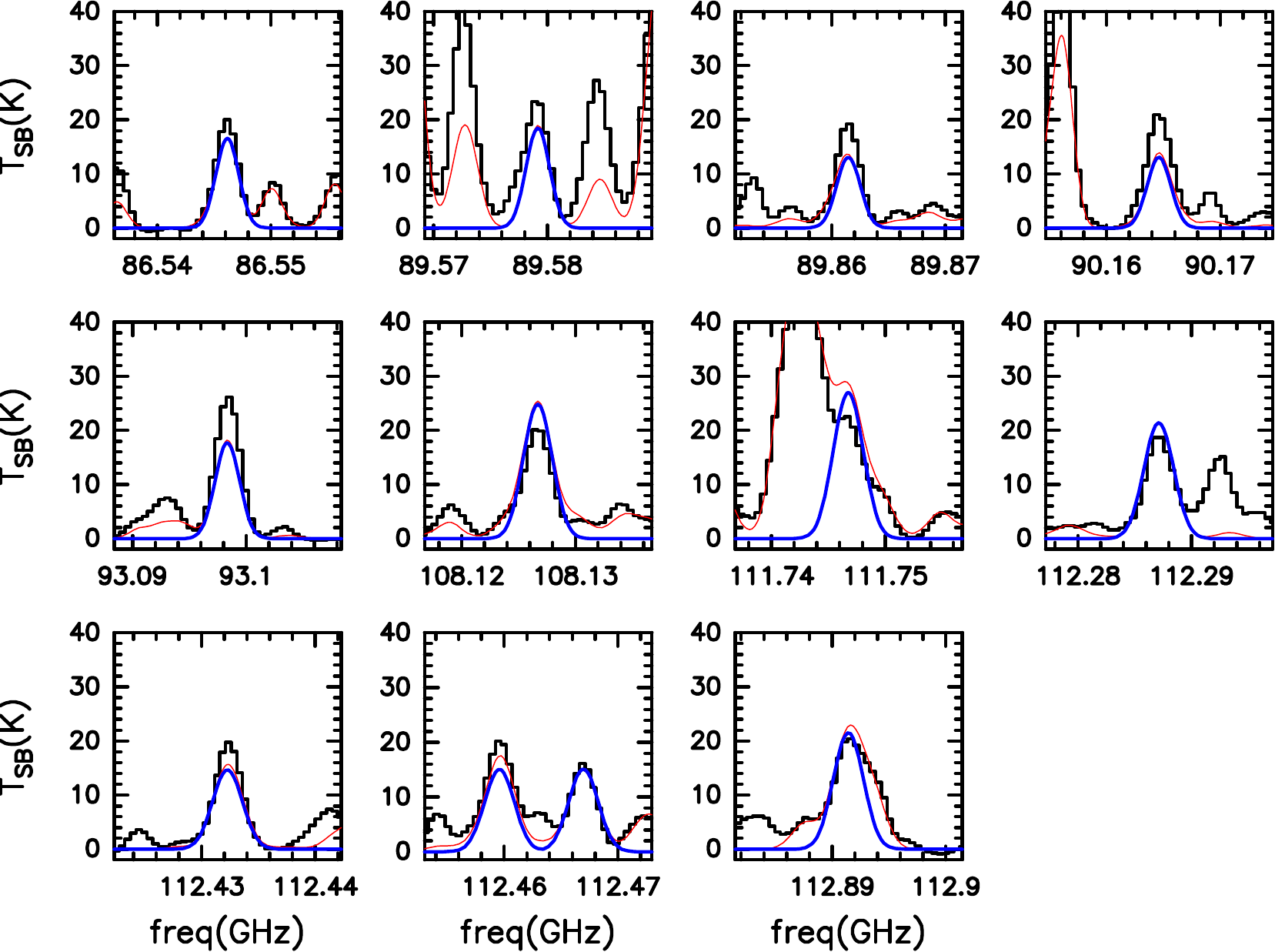}
\caption{Transitions listed in Table~\ref{transitions} and used to fit the trans isomer of HCOOH (t-HCOOH). The blue curve represents the best LTE fit obtained with MADCUBA and the red curve shows the simulated spectrum taking into account all the species identified in the region so far.}
\label{fig-t-formic}
\end{figure*}

\begin{figure*}
\centering
\includegraphics[width=35pc]{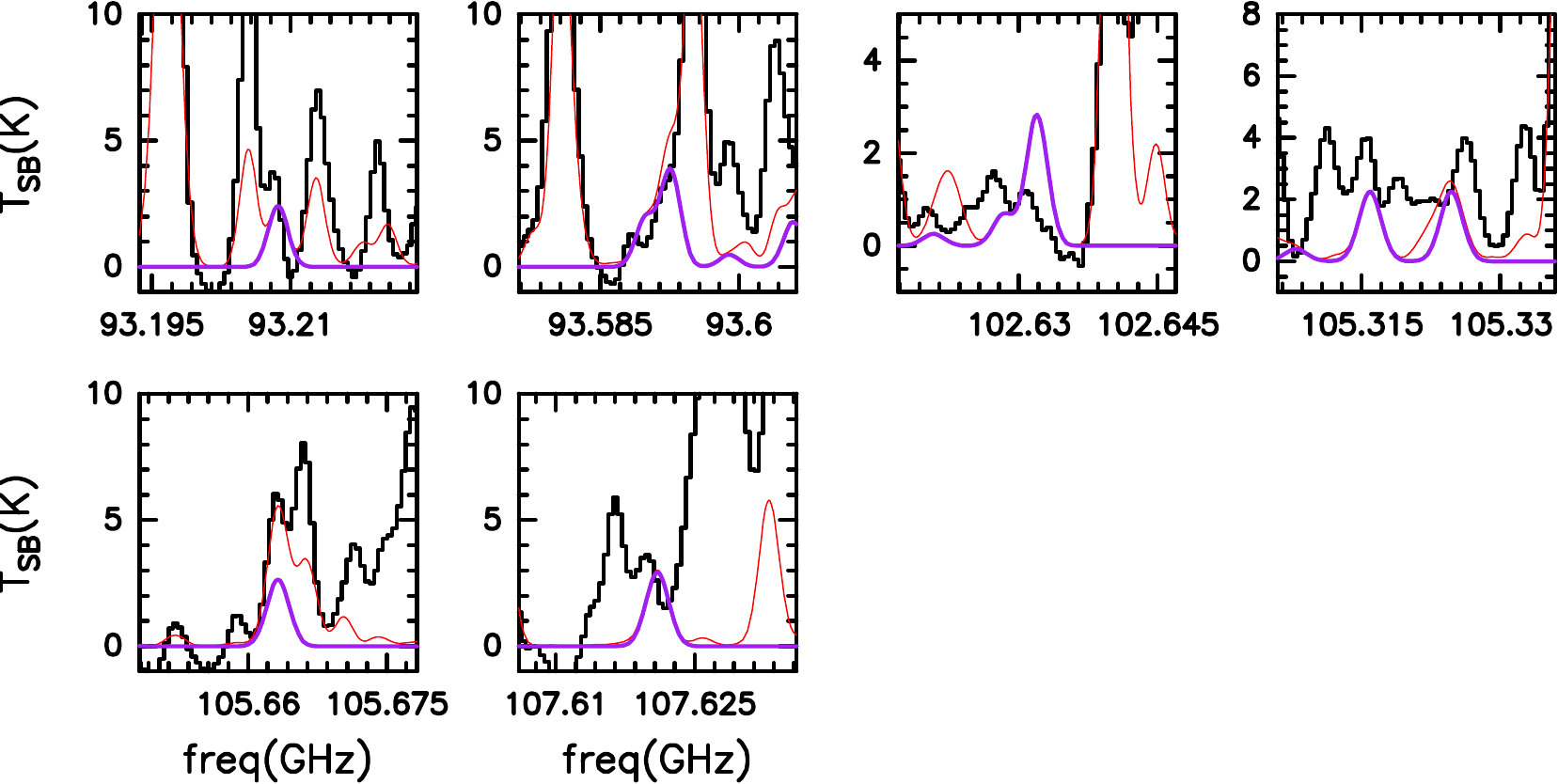}
\caption{Transitions listed in Table~\ref{transitions} and used to fit the trans form of HC(O)SH (t-HC(O)SH). The purple curve represents the best LTE fit obtained with MADCUBA and the red curve shows the simulated spectrum taking into account all the species identified in the region so far. We note that the discrepancy between the predicted and observed spectra at 102.63 GHz may be due to a poor baseline removal at this frequency range and/or to some absorption associated with the strong molecular line centered at 102.645 GHz.}
\label{fig-t-tio}
\end{figure*}

The GUAPOS spectral survey was carried out with ALMA during Cycle 5 (project 2017.1.00501.S, P.I.: M. T. Beltrán) in band 3 with a frequency coverage from 84.05 GHz up to 115.91 GHz. The frequency resolution of the observation was 0.49 MHz, corresponding to a velocity resolution of $\sim$1.6 km s$^{-1}$ at 90 GHz. The final angular resolution of the observations was $\sim$1\farcs2 ($\sim$4500 au). The central coordinates were $\alpha_{\rm J2000}$ = 18$^{\rm h}$47$^{\rm m}$34$^{\rm s}$ and $\delta_{\rm J2000}$ = $-$0.1\degree12$^{\prime}$45\asec. The uncertainties in the flux calibration were of $\sim$5\%. An additional error of 11\% should also be considered due to the determination of the continuum level. For more details see \citet{mininni2020} and \citet{laura21}.

\begin{figure*}
\centering
\includegraphics[width=35pc]{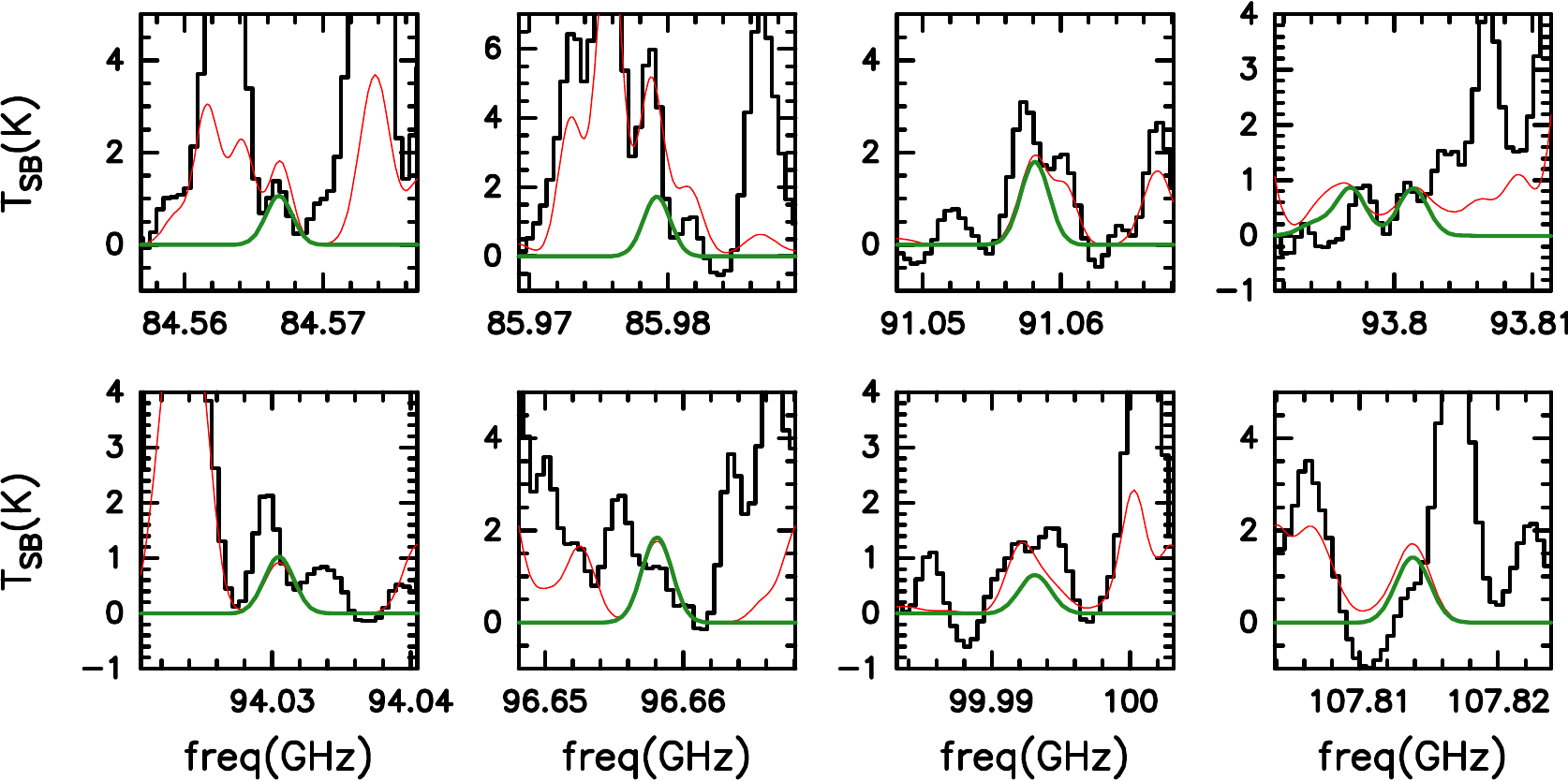}
\caption{Transitions listed in Table~\ref{transitions} and used to fit the cis form of HC(O)SH (c-HC(O)SH). The green curve represents the best LTE fit obtained with MADCUBA and the red curve shows the simulated spectrum taking into account all the species identified in the region so far.}
\label{fig-c-tio}
\end{figure*}

The identification of the transitions of the cis and trans isomers of HCOOH and HC(O)SH in the GUAPOS spectrum was performed using version 01/12/2020 of the SLIM (Spectral Line Identification and Modeling) tool within the MADCUBA package\footnote{Madrid Data Cube Analysis on ImageJ is a software developed at the Center of Astrobiology (CAB) in Madrid; \url{http://cab.intacsic.es/madcuba/Portada.html}.} (\citealt{martin2019}). SLIM uses the spectroscopic entries from the Cologne Database for Molecular Spectroscopy\footnote{\url{http://cdms.astro.uni-koeln.de/classic/}.}(CDMS, \citealt{muller2001}; \citealt{muller2005}; \citealt{endres2016}) and the Jet Propulsion Laboratory\footnote{\url{https://spec.jpl.nasa.gov/ftp/pub/catalog/catdir.html}.} (JPL, \citealt{pickett1998}), and generates a synthetic spectrum, assuming local thermodynamic equilibrium (LTE) conditions and taking into account the line opacity. For each molecular species, we used the MADCUBA-AUTOFIT tool to compare the observed spectrum with the LTE synthetic one. This tool takes into account four physical parameters to create a synthetic LTE line profile: the column density ($N$), the excitation temperature ($T_{\rm ex}$), peak radial velocity (v$_{\rm LSR}$), and the full width half maximum ($FWHM$). The LTE assumption is well justified given the high volume density of the source $n$(H$_{2}$) $\sim$ 10$^{8}$ cm$^{-3}$ (\citealt{mininni2020}). The parameters are left free to obtain the best-LTE fit. When convergence is not reached, the parameters $FWHM$, v$_{\rm LSR}$, and/or $T_{\rm ex}$ are fixed. The best-LTE fit of the GUAPOS spectrum considering all molecular species analyzed so far within the survey can be found in Appendix E of \citet{laura21}.

The spectroscopy of t-HCOOH and c-HCOOH is taken from \citet{winnewisser2002} and their dipole moments were measured by \citet{kuze1982} for t-HCOOH and \citet{hocking1976a} for c-HCOOH. The non-contaminated transitions used to fit the molecular lines of t-HCOOH within SLIM are listed in Table \ref{transitions}. Unfortunately, non-contaminated transitions of c-HCOOH have not been found.

Figure \ref{fig-t-formic} shows the non-contaminated transitions used to fit t-HCOOH. To perform the fit we fixed the v$_{\rm LSR}$ to 96.5 km s$^{-1}$, which is the velocity of the source. The best-fit parameters obtained with MADCUBA are $T_{\rm ex}$ = 152$\pm$45 K, $FWHM$ = 8.0$\pm$0.3 km s$^{-1}$, and $N$ = (1.4$\pm$0.6)$\times$10$^{17}$ cm$^{-2}$. This implies an abundance of $X$ = (1.4$\pm$0.7)$\times$10$^{-8}$ for the H$_2$ column density of $N_{\rm H_{2}}$ = (1.0 $\pm$ 0.2) $\times$10$^{25}$ cm$^{-2}$ measured by \citealt{mininni2020} at 3mm. On the other hand, c-HCOOH has not been detected and hence we derived a rough estimate of the upper limit of the column density using MADCUBA-SLIM, given the difficulty of obtaining a precise result due to the blending with lines of other species. Thus, we assumed $T_{\rm ex}$, v$_{\rm LSR}$ and $FWHM$ derived for t-HCOOH, and increased $N$ to the maximum value compatible with the observed spectrum. Finally, we obtained $N$ $\le$1.6$\times$10$^{15}$ cm$^{-2}$. This column density translates into an upper limit to the molecular abundance of c-HCOOH of $X\le$1.6$\times$10$^{-10}$. The trans/cis ratio measured toward G31 is thus $\ge$90. We note that if we use other $T_{\rm ex}$ ranging from 50 up to 300$\,$K, we still get a trans/cis lower limit between $\geq$70 and $\geq$95. 

\section{Discussion}
\subsection{Isomerization of formic acid, HCOOH, in the ISM}
\label{formic}

In Table \ref{table-t-c-ratio}, we compare our G31 results with the trans/cis abundance ratio measured toward other astronomical sources. These sources range from cold cores such as B5 and L483 \citep[with temperatures $\sim$5-15$\,$K;][]{taquet17,agundez19}, to massive hot cores \citep[as e.g., G31 and W51 e2, with temperatures $\sim$100-200$\,$K; see this work and][]{rivilla2017b}, and the quiescent giant molecular cloud, G+0.693-0.027, located in the Galactic center \citep[with temperatures between 70 and 150$\,$K; see][]{lucas21}.  From this table, we find that the trans/cis ratios measured toward massive hot cores and the G+0.693-0.027 Galactic center molecular cloud are consistent with the small amount of c-HCOOH expected under thermodynamic equilibrium (Table\ref{tab1}), which makes it undetectable even at high temperatures.

However, at low temperatures, the predicted trans/cis ratios  differ from those measured in the cold molecular cores B5 and L483
---by multiple orders of magnitude--- where the trans/cis ratios found for HCOOH are 15 and 17, respectively. Indeed, the isomerization of HCOOH acid should not occur at these low temperatures (see the low values for the $K_{1}$ and $K_{-1}$ rate constant in Table 1 for T$\leq$30 K). This implies that the ratios measured between the isomers of HCOOH at low temperatures must be due to competitive chemical processes and are not related to an isomerization transformation.

\begin{table*}
\caption{Unblended or slightly blended transitions of the molecules studied in this work. The first column indicates the molecule for which the transitions are listed. The second column shows the frequency of the rotational transition given in the third column. log$I$ is the base 10 logarithm of the integrated intensity at 300 K, $E_{\rm up}$ is the energy of the upper level and $\tau$ is the optical depth of the transition derived from the LTE fit, as described in Sect.~\ref{observation}.}             
\label{transitions}      
\centering           
\begin{tabular}{l c c  c c c}  
\hline\hline    
Molecule & Frequency    & Transition    & log$I$        & $E_{\rm up}$          &$\tau$ \\
& (MHz) & ($J$, $K_{\rm a}$, $K_{\rm c}$)  & (nm$^{2}$ MHz) &  (K) & \\
\hline                        
t-HCOOH & 86546.1891 & 4,1,4--3,1,3 & -4.3888 & 13.57 & 0.12$\pm$0.08 \\
t-HCOOH & 89579.1785 & 4,0,4--3,0,3 & -4.3269 & 10.77 & 0.13$\pm$0.09 \\
t-HCOOH & 89861.4843 & 4,2,3--3,2,2 & -4.4673 & 23.51 & 0.09$\pm$0.06 \\
t-HCOOH & 90164.6296 & 4,2,2--3,2,1 & -4.4644 & 23.53 & 0.09$\pm$0.06 \\
t-HCOOH & 93098.3627 & 4,1,3--3,1,2 & -4.3263 & 14.36 & 0.13$\pm$0.09 \\
t-HCOOH & 108126.7202 & 5,1,5--4,1,4 & -4.095 & 18.76 & 0.18$\pm$0.12 \\
t-HCOOH & 111746.7845 & 5,0,5--4,0,4 & -4.0451 & 16.13 & 0.20$\pm$0.14 \\
t-HCOOH & 112287.1447 & 5,2,4--4,2,3 & -4.1347 & 28.90 & 0.16$\pm$0.10 \\
t-HCOOH & 112432.2921 & 5,4,2--4,4,1 & -4.5567 & 67.06 & 0.05$\pm$0.03 \\
t-HCOOH & 112432.3191 & 5,4,1--4,4,0 & -4.5567 & 67.06 & 0.05$\pm$0.03 \\
t-HCOOH & 112459.6214 & 5,3,3--4,3,2 & -4.2744 & 44.81 & 0.11$\pm$0.07 \\
t-HCOOH & 112467.0074 & 5,3,2--4,3,1 & -4.2744 & 44.81 & 0.11$\pm$0.07 \\
t-HCOOH & 112891.4435 & 5,2,3--4,2,2 & -4.1301 & 28.95 & 0.16$\pm$0.10 \\
\hline
t-HC(O)SH & 93208.624 & 8,0,8--7,0,7 & -4.3707 & 20.17 &  0.016 $\pm$0.003 \\
t-HC(O)SH & 93589.887 & 8,5,4--7,5,3 & -4.6797 & 87.60 &  0.0065$\pm$0.0011 \\
t-HC(O)SH & 93589.887 & 8,5,3--7,5,2 & -4.6797 & 87.60 &  0.0065$\pm$0.0011 \\
t-HC(O)SH & 93592.509 & 8,4,5--7,4,4 & -4.5544 & 63.35 &  0.0093$\pm$0.0016 \\
t-HC(O)SH & 93592.509 & 8,4,4--7,4,3 & -4.5544 & 63.35 &  0.0093$\pm$0.0016\\
t-HC(O)SH & 93592.808 & 8,6,2--7,6,1 & -4.8664 & 117.22 &  0.0038$\pm$0.0007 \\
t-HC(O)SH & 93592.808 & 8,6,3--7,6,2 & -4.8664 & 117.22 &  0.0038$\pm$0.0007 \\
t-HC(O)SH & 102631.9508 & 9,1,9--8,1,8 & -4.2512 & 27.34 &  0.019$\pm$0.003 \\
t-HC(O)SH & 105315.909 & 9,3,7--8,3,6 & -4.3065 & 49.54 &  0.015$\pm$0.003 \\
t-HC(O)SH & 105324.697 & 9,3,6--8,3,5 & -4.3065 & 49.54 &  0.015$\pm$0.003 \\
t-HC(O)SH & 105663.246 & 9,2,7--8,2,6 & -4.2551 & 36.10 &  0.018$\pm$0.003 \\
t-HC(O)SH & 107620.9874 & 9,1,8--8,1,7 & -4.2115 & 28.54 &  0.020$\pm$0.003 \\
\hline
c-HC(O)SH & 84566.811 & 11,0,11--10,1,10 & -4.1819 & 36.99 & 0.0072$\pm$0.0018 \\
c-HC(O)SH & 85979.209 & 13,1,12--13,0,13 & -3.9382 & 55.05 & 0.012$\pm$0.003 \\
c-HC(O)SH & 91058.108 & 14,1,13--14,0,14 & -3.8857 & 63.09 & 0.012$\pm$0.003 \\
c-HC(O)SH & 93801.341 & 8,3,5--7,3,4 & -4.2206 & 44.91 & 0.0058$\pm$0.0015\\
c-HC(O)SH & 94030.48 & 8,2,6--7,2,5 & -4.1609 & 31.24 & 0.0069$\pm$0.0017 \\
c-HC(O)SH & 96658.0938 & 15,1,14--15,0,15 & -3.8353 & 71.69 & 0.012$\pm$0.003 \\
c-HC(O)SH & 99993.1226 & 19,1,18--18,2,17 & -4.189 & 111.77 & 0.0047$\pm$0.0012 \\
c-HC(O)SH & 107813.858 & 9,1,8--8,1,7 & -3.9641 & 28.63 & 0.010$\pm$0.002 \\
\hline                                   
\end{tabular}
\end{table*}

\begin{table*}
\setlength{\tabcolsep}{5pt}
\caption{Trans/cis ratio for HCOOH derived in the ISM compared with those predicted by this work for isomerization reactions.}
\centering
  \begin{tabular}{lcccc}
  \hline
    Source & $T_{\rm kin}^{a}$ & Observed trans/cis & Predicted trans/cis & Reference\\ 
         & (K) &  & ($K{_1}$/$K_{-1}$ at $T_{\rm kin}$) & \\ 
  \hline
  Barnard 5 & 5--15 & 15 &$>$10$^{50}$ & (1) \\
  L483 & 10 &17 & 2$\times$10$^{88}$&(2) \\
  G31 & 150$^{b}$ & $\ge$90 & 8$\times$10$^{5}$ & (3)\\
  G+0.693-0.027 & 150$^c$ & $\ge$125& 8$\times$10$^{5}$ & (4) \\
  W51 e2 & 90--180 & $\ge$16 & 7$\times$10$^{9}$--8$\times$10$^{4}$ & (5)\\
  \hline
  \normalsize
  \label{table-t-c-ratio}
  \end{tabular}
   \newline 
  \textbf{References.}
(1) \citet{taquet17}; (2) \citet{agundez19}; (3) This work; (4) \citet{lucas21}; (5) \citet{rivilla2017b}. 
\newline
  \textbf{Notes.} $^{a}$ $T_{\rm kin}$ is the gas kinetic temperature of the source inferred from t-HCOOH. ${^b}$ This is an average kinetic temperature towards the G31 hot molecular core (e.g., \citealt{beltran2018}). $^c$ Inferred from CH$_3$CN data by \citet{zeng2018}.
      \normalsize
\end{table*}

\begin{table*}
\setlength{\tabcolsep}{5pt}
\caption{Trans/cis ratio for HC(O)SH derived in the ISM compared with those predicted by this work for isomerization reactions.}
\centering
  \begin{tabular}{lcccc}
  \hline
    Source & $T_{\rm kin}^{a}$& Observed trans/cis & Predicted trans/cis & Reference\\ 
         & (K) &  & ($K{_1}$/$K_{-1}$ at $T_{\rm kin}$)& \\ 
  \hline
  G31 & 150$^b$ & 3.7$\pm$1.5$^{c}$ & 8.6 & (1)\\
  G+0.693-0.027 & 150$^d$ & $\ge$5 & 8.6 & (2) \\
  \hline
  \normalsize
  \label{table-t-c-ratio2}
  \end{tabular}
   \newline 
  \textbf{References.} (1) This work; (2) \citet{lucas21}.
   \newline 
\textbf{Notes.} $^{a}$ $T_{\rm kin}$ is the gas kinetic temperature of the source inferred from t-HCOOH. $^{b}$ This is an average kinetic temperature towards the G31 hot molecular core (e.g., \citealt{beltran2018}). $^{c}$This ratio is obtained from two tentative detections and should be taken with caution. $^d$ Inferred from CH$_3$CN data by \citet{zeng2018}.
      \normalsize
\end{table*}

We note that we do not compare our results with the trans/cis ratios measured toward the Orion Bar because the production of c-HCOOH in this source is due to the gas being irradiated by UV photons (\citealt{cuadrado16}). This process enables photoisomerization processes by exciting molecules to higher electronic states, which is a completely different mechanism to that of the isomerization transformation through tunneling.

\subsection{Isomerization of thioformic acid, HC(O)SH, in the ISM}
\label{thioformic}

As mentioned in the introduction, HC(O)SH acid has recently been reported in the ISM toward the Galactic center giant molecular cloud G+0.693-0.027 \citep{lucas21}. In this section, we also search for this species in the G31 massive hot core making use of the GUAPOS spectral survey. For this, we used the spectroscopic information of t-HC(O)SH and c-HC(O)SH obtained by \citet{hocking1976b} and the dipole moments measured by \citet{hocking1976c}. The noncontaminated transitions used to fit the molecular lines of t-HC(O)SH and c-HC(O)SH can also be found in Table \ref{transitions}.

The trans and cis forms of HC(O)SH are tentatively detected towards G31. Figures~\ref{fig-t-tio} and \ref{fig-c-tio} report the less contaminated transitions of these species for which we derive a tentative $N$. For t-HC(O)SH the main contaminants are acetone (CH$_{3}$COCH$_{3}$) and ethylene glycol (HOCH$_{2}$CH$_{2}$OH) at 105.663 GHz and acetaldehyde (CH$_{3}$CHO) at 93.595 GHz. For these two lines, the total observed emission can be perfectly fitted taking into account the emission from t-HC(O)SH. The observed spectra at the frequencies of the remaining lines cannot be completely reproduced due to the contribution from unidentified species. The main contaminants for c-HC(O)SH are CH$_{3}$OCH$_{3}$ at 85.978 GHz, CH$_{3}$COOH at 85.981 GHz, CH$_{3}$CONH$_{2}$ at 84.567 GHz, and CH$_{3}$OCHO at 99.992 GHz. We fixed $T_{\rm ex}$, v$_{\rm LSR}$, and $FWHM$ to those inferred for t-HCOOH and we obtained a $N$ = (2.0$\pm$0.5)$\times$10$^{16}$ cm$^{-2}$ and an abundance of $X$ = (2.0$\pm$0.6)$\times$10$^{-9}$ for t-HC(O)SH, and a column density of $N$ = (5.4$\pm$1.5)$\times$10$^{15}$ cm$^{-2}$ and an abundance of $X$ = (5.4$\pm$1.8)$\times$10$^{-10}$ for c-HC(O)SH. The resulting trans/cis ratio for HC(O)SH acid is 3.7$\pm$1.5. We remind the reader that this value has been inferred from the ratio of two tentative detections and should be taken with caution.

In Table \ref{table-t-c-ratio2}, we compare the G31 results with those obtained for G+0.693-0.027 and our kinetic results of Section 3. From this table, we find that the ratio of the rate constants between the forward (cis to trans) and backward (trans to cis) reactions are consistent with the trans/cis ratios measured toward both sources. In fact, assuming a kinetic temperature $\geq$240$\,$K for G31 (\citealt{beltran2018}), the predicted $K{_1}$/$K_{-1}$ (trans/cis) ratios fall below $\sim$4, which nicely matches the tentative trans/cis ratio measured toward G31 with the GUAPOS data. Therefore, the comparison of our kinetic results with observations suggests that c-HC(O)SH is likely present in hot environments where the isomerization transformation due to tunneling can take place.

\section{Conclusions}

In this work we carried out a theoretical analysis of the cis-to-trans and trans-to-cis isomerization reactions of the HCOOH acid and of the recently detected HC(O)SH thioacid. The kinetic results show that the trans/cis ratio found in the ISM for both acids could follow a behaviour typical of the thermodynamic equilibrium, independently of the formation route, thanks to ground-state quantum tunneling effects. 

Our kinetic results reveal that, at very low temperatures, the isomerization reactions cannot occur because their rate constants are very low. This is in disagreement with the observed trans/cis ratios of HCOOH acid in the cold molecular cores B5 and L483, which implies that chemical processes such as competitive chemical routes or excitations caused by secondary UV fields must be responsible.

We also searched for these molecules in the context of the GUAPOS spectral survey obtained with the ALMA interferometer towards the hot core G31. We detect the trans form of HCOOH and tentatively detected trans and cis HC(O)SH, the latter for the first time in the ISM. For the column density of c-HCOOH, we obtained an upper limit. At high temperatures, our theoretical predictions support the upper limits found for HCOOH acid in G31 (trans/cis \(\geq\)90), G+0.693-0.027 (trans/cis \(\geq\)125), and W51 e2 (trans/cis \(\geq\)16). The energy difference between the c-HCOOH and the more stable t-HCOOH causes the equilibrium to displace toward the latter. This phenomenon means that the amount of the c-HCOOH in the ISM is very low, and even undetectable in hot molecular clouds. For HC(O)SH acid, our tentative detections of both isomers toward G31 within the GUAPOS data provides a trans/cis ratio of 3.7$\pm$1.5, which is close to the ratio of the equilibrium constants computed at $\geq$240 K ($\leq$4). Our quantum chemical calculations also support the upper limit measured for the trans/cis abundance ratio of HC(O)SH acid toward the G+0.693-0.027 molecular cloud (\(\geq\)5).

\begin{acknowledgements}

J.G.d.l.C. acknowledges the Spanish State Research Agency (AEI) through project number MDM-2017-0737 Unidad de Excelencia “María de Maeztu”—Centro de Astrobiología
(CSIC-INTA). J.C.C. acknowledges the Junta de Extremadura and European Regional Development Fund, Spain, Project No. GR18010. I.J.-S. and J.M.-P. have received partial support from AEI (PID2019-105552RB-C41). G. M. acknowledges the support of the Alexander von Humboldt Foundation through a postdoctoral research grant. V.M.R. and L.C have received funding from the Comunidad de Madrid through the Atracción de Talento Investigador (Doctores con experiencia) Grant (COOL: Cosmic Origins Of Life; 2019-T1/TIC-15379). This paper makes use of the following ALMA data: ADS/JAO.ALMA\#2017.1.00501.S. ALMA is a partnership of ESO (representing its member states), NSF (USA) and NINS (Japan), together with NRC (Canada), MOST and ASIAA (Taiwan), and KASI (Republic of Korea), in cooperation with the Republic of Chile. The Joint ALMA Observatory is operated by ESO, AUI/NRAO and NAOJ. Computational assistance was provided by the Supercomputer facilities of LUSITANIA founded by Cénits and Computaex Foundation.

\end{acknowledgements}

\bibliographystyle{aa}
\bibliography{bibliography}

\begin{appendix}
\section{Comparison Between Coupled Cluster and Double-Hybrid Calculations}

Table A.1 shows the relative electronic energies of the cis, trans, and saddle point for the isomerization of the HCOOH relative to the cis isomer. These results are obtained after a full geometry optimization with the selected methods. To characterize the nature of the stationary points, frequency calculations were also carried out at the same level of theory. Thus, for comparison, the harmonic zero-point energies are
also given. The great agreement obtained between both methods indicates that this double hybrid is an excellent method for quantitative studies for this system. This agreement in the energies is a reflection of the similarity in the optimized geometries between both methods, which give almost the same results and are gathered in the Table A.2. As a complement of Table A.2, the Figure A.1 shows the atom numbering of the formic acid used in this table.

\begin{table}[h!]
\caption{Relative electronic energies and harmonic ZPE of the cis, trans, and TS optimized geometries of HCOOH at CCSD(T)-F12 and B2PLYP-D3(BJ) level of theory}             
\label{tabA1}      
\centering           
\begin{tabular}{c c c c c}  
\hline\hline    
 & \multicolumn{2}{c}{B2PLYP-D3(BJ)$^{a}$} & \multicolumn{2}{c}{CCSD(T)-F12$^{b}$}\\
\hline
 & \(\Delta\)$E$ & $ZPE_{har}$ & \(\Delta\)$E$ & $ZPE_{har}$\\
\hline                        
cis     &   0.00        &       20.96   &       0.00    &       20.29   \\
TS      &   8.77        &       19.90   &       8.39    &       20.02   \\
trans   &  -4.15        &       21.18   &  -4.25        &       21.29   \\
\hline                                   
\end{tabular}
\\
\textbf{Notes.} $^{a}$ In combination with the aug-cc-pVTZ basis set. $^{b}$ In combination with the cc-pVTZ-F12-CABS and cc-pVTZ-F12 basis set and the augmented aug-cc-pVTZ for correlation and coulomb fitting.
\end{table}

\begin{figure}[h!]
  \centering
  \includegraphics[width=5cm,height=4.5cm]{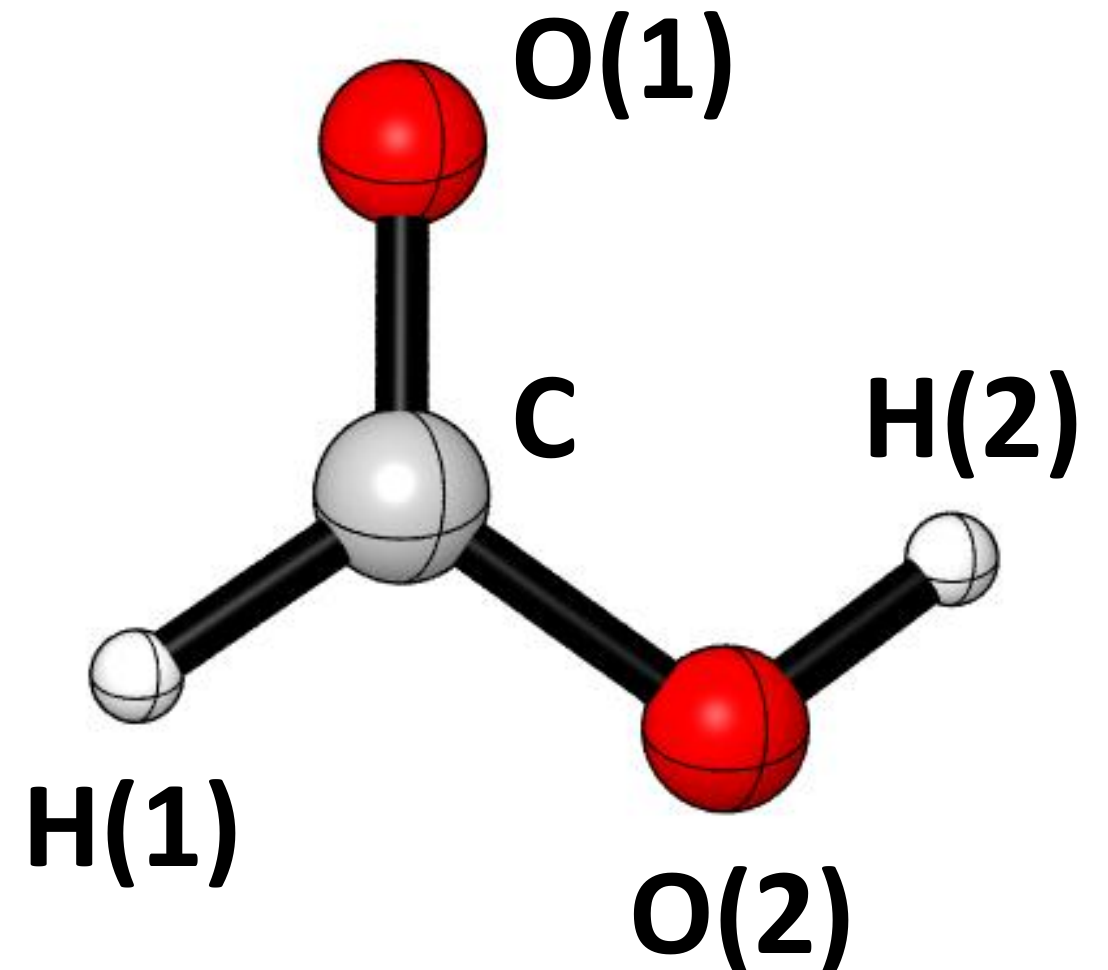}
  \caption{Atom numbering of HCOOH.}
\end{figure}

\begin{table*}
\setlength{\tabcolsep}{3pt}
\caption{Geometrial parameters of the cis, trans, and TS optimized geometries of HCOOH at CCSD(T)-F12 and B2PLYP-D3(BJ) level of theory. Bond distances are given in angstroms and angles in degrees.}             
\label{tabA2}      
\centering           
\begin{tabular}{c c c c c c c}  
\hline\hline    
 & \multicolumn{2}{c}{cis} & \multicolumn{2}{c}{TS} & \multicolumn{2}{c}{trans} \\
\hline
 & B2PLYP-D3(BJ)$^{a}$ & CCSD(T)-F12$^{b}$ & B2PLYP-D3(BJ)$^{a}$ & CCSD(T)-F12$^{b}$ & B2PLYP-D3(BJ)$^{a}$ & CCSD(T)-F12$^{b}$ \\
\hline                        
C-O(1)  &       1.19    &       1.19    &       1.19    &       1.19    &       1.20    &       1.20    \\
C-O(2)  &       1.35    &       1.35    &       1.38    &       1.37    &       1.35    &       1.34    \\
C-H(1)  &       1.10    &       1.10    &       1.10    &       1.10    &       1.09    &       1.09    \\
O(2)-H(2)       &       0.96    &       0.96    &       0.96    &       0.96    &       0.97    &       0.97    \\
O(1)-C-O(2)     &       122.40  &       122.30  &       123.77  &       123.60  &       125.07  &       124.91  \\
C-O(2)-H(2)     &       109.70  &       109.21  &       111.49  &       110.57  &       107.30  &       106.74  \\
H(1)-C-O(2)     &       113.53  &       113.70  &       112.51  &       112.80  &       109.73  &       109.98  \\
H(1)-C-O(1)     &       124.07  &       124.00  &       123.60  &       123.49  &       125.20  &       125.11  \\
H(2)-O(2)-C-H(1)        &       0.00    &       0.00    &       -89.64  &       -89.29  &       179.99  &       180.00  \\
H(2)-O(2)-C-O(1)        &       180.00  &       180.00  &       94.27   &       94.33   &       0.01    &       0.00    \\
\hline                                   
\end{tabular}
\\
\textbf{Notes.} $^{a}$ In combination with the aug-cc-pVTZ basis set. $^{b}$ In combination with the cc-pVTZ-F12-CABS and cc-pVTZ-F12 basis set and the augmented aug-cc-pVTZ for correlation and coulomb fitting.
\end{table*}

\end{appendix}

\end{document}